# Lithium niobate-enhanced laser photoacoustic spectroscopy

Haoyang Lin, Wenguo Zhu, Yongchun Zhong, Jieyuan Tang, Huihui Lu, Jianhui Yu, and Huadan Zheng*

*Department of Optoelectronic Engineering, Jinan University, Guangzhou, 510632, China*
*\*Email correspondance : zhenghuadan@jnu.edu.cn*

In this paper, the photoacoustic spectroscopy technique based on lithium niobate crystals is initially reported, to our knowledge. A novel dual-cantilever tuning fork structure and new electrodes have been designed using Y-cut 128° blackened lithium niobate wafers. The tuning fork, with a resonant frequency of only 10.46 kHz and a prong gap of 1 mm, is engineered to achieve superior performance in photoacoustic spectroscopy. In the demonstration experiment, acetylene was detected using a 1.53 um semiconductor laser, achieving a detection limit of about 9 ppb within a one-second integration time.

Trace gas detection is pivotal across various domains, serving critical functions in environmental monitoring, industrial process, healthcare, agriculture, food safety, and national security [1-5]. Photoacoustic spectroscopy (PAS), was used extensively in various gas detection applications [6,7]. By PAS, high sensitivity can be achieved in a small volume of gas [8].

Due to its unique double-cantilever beam structure and vibration mode, the tuning fork (TF) sensor can achieve both stable frequency and high-quality factor, offering simplicity in production, lower cost, and ease of implementation. Quartz tuning forks (QTFs), which were among the earliest researched, are now widely utilized in a variety of applications. Their applications span across high-resolution mechanical sensors, vibration amplitude sensors for vibrating magnetic samples, fluid sensor, and trace gas detection [9-12].

The stability of lithium niobate (LN) crystals at high temperatures, evidenced by their Curie temperature of 1210 °C, surpasses that of quartz crystals, which have a Curie temperature of only 575 °C. The piezoelectric coefficient of LN-TF shows a significant improvement compared to that of QTF. LN forks utilize $d23$, with a peak of up to $28\times10^{-12}$ C/N, in contrast to most QTF that use $d21$, peaking at $6\times10^{-12}$ C/N [9]. The electromechanical coupling coefficient of LN crystals is significantly higher, reaching 0.68, compared to 0.3 for quartz crystals [10]. This high electromechanical coupling coefficient, along with their ability to withstand high temperatures and pressures, has made LN crystals increasingly favored for piezoelectric transducers and sensors [11].

Figure 1 displays the charge distribution of lithium niobate (LN) crystal tuning forks in various cutting orientations, simulated using the COMSOL software. It is important to note that, although the tuning fork resonators possess multiple vibration modes, only the symmetric vibration model is considered in this context. Future work will discuss other vibration modes. Figures 1(a), (b), and (c) illustrate the charge distributions for tuning forks based on X-cut, Y-cut, and Z-cut orientations, respectively. In each orientation, there are two different directions for the tuning fork tines, as indicated by the XYZ coordinate axes in the diagram. Despite having the same crystal cut orientation, the tines of the forks differ by a 90° orientation. Notably, in the Y-cut orientation, the charge predominantly accumulates on the surface of the fork's tines rather than on the sides. Such a distribution of charge is advantageous for maximizing charge collection.

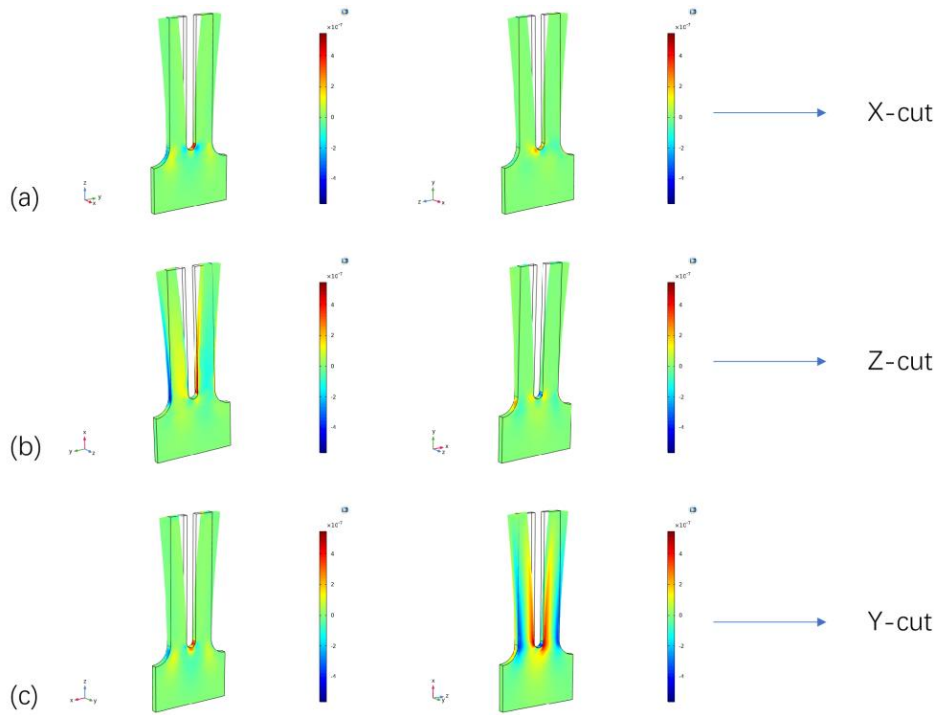

Figure 1 exhibits tuning forks fabricated from lithium niobate crystals in x-cut, y-cut, and z-cut orientations.

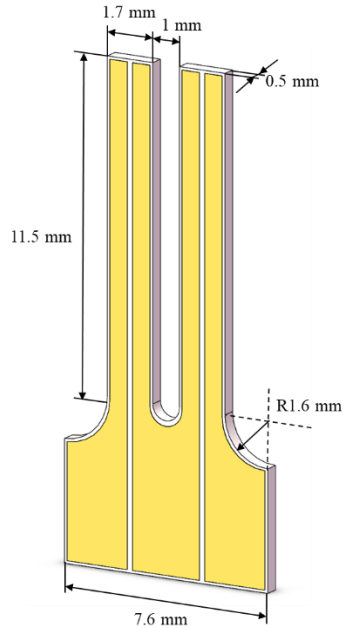

Figure 2 illustrates the structure and electrode design of a tuning fork based on Y-cut 128° lithium niobate crystal.

Upon reviewing the existing literature, it is identified that the Y-cut 128° direction is a comparatively ideal orientation. Hence, in this work, the Y-cut 128° orientation is selected, bearing a charge distribution similar to that of the standard Y-cut. A lithium niobate tuning fork, as depicted in Figure 2, has been fabricated. A commercial lithium niobate wafer with a thickness of 0.5 mm was chosen. This wafer is blackened to reduce the pyroelectric effect inherent in lithium niobate, thereby lowering the background noise during fork detection. The tuning fork's tines have a length of 11.5 mm, a width of 1.7 mm, and a gap of 1 mm between them. The outer sides of the tines are chamfered in a circular manner. This tine gap facilitates easier collimation when using infrared light sources, terahertz sources, high-power light sources, and incoherent sources such as LEDs, thus reducing scattered light noise.

The electrodes, as shown in Figure 2, are fabricated using a sputtering technique. These electrodes divide each face of the tuning fork into three parts: the middle part collects one positive charge, while the symmetric outer parts jointly collect another negative charge. The backside of the fork features an identical electrode distribution, but with reversed charge polarity. This novel electrode fabrication method is relatively simple but highly effective in collecting charges.

In photoacoustic spectroscopy, an acoustic resonator is a vital method to enhance the sound wave coupling efficiency of TF transducers. Therefore, an acoustic micro resonator (AmR) was fabricated and adapted to the lithium niobate tuning fork. Composed of stainless-steel capillaries, the resonator is configured in a coaxial arrangement, allowing the formation of one-dimensional acoustic standing waves, thereby enhancing the fork's resonant intensity. The resonance curve of the lithium niobate tuning fork was measured, both before and after the integration of the AmR, as shown in Figure 3. Following a Lorentzian fit, the resonance frequency of the lithium niobate tuning fork was found to be 10461.7 Hz with a quality factor (Q value) of 1439. After coupling with the AmR, the resonance frequency slightly shifted to 10460 Hz, and the quality factor decreased to 1295. This reduction in the quality factor post-coupling is attributed to the transfer of sound wave energy. Further optimization of the tuning fork's parameters and fabrication process could potentially elevate the quality factor to 8000-10000.

A semiconductor laser with a wavelength of 1.53 μm, in conjunction with an optical fiber EDFA, was employed to detect acetylene gas molecules, thereby verifying the enhanced photoacoustic spectroscopy detection performance of the lithium niobate tuning fork. Consultation of the HITRAN database led to the selection of an absorption line at 6507.39 cm$^{-1}$, with an absorption strength of $4.88 \times 10^{-21}$ cm/molecule. The output wavelength of the semiconductor laser was tuned by adjusting its injection current, covering the selected acetylene molecular absorption line. Figure 4 shows the demodulated second-harmonic signal. Standard mixed gas with 50 ppm of acetylene yielded a detection signal-to-noise ratio of 5200, achieving a detection limit of ~9 ppb.

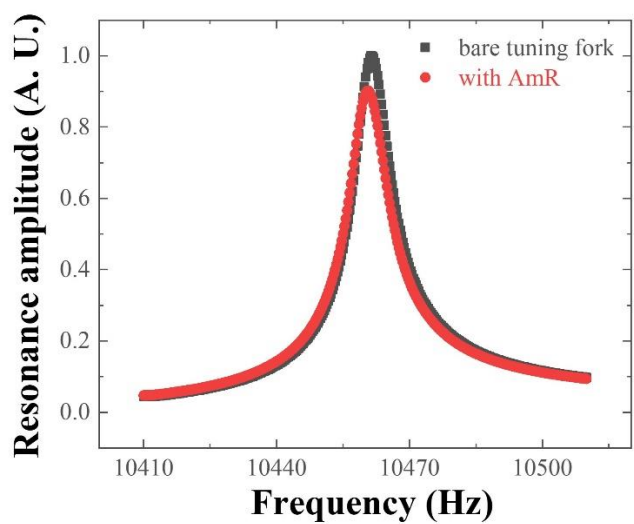

Figure 3 displays the resonance curve of the lithium niobate tuning fork and the curve after coupling with an acoustic resonator.

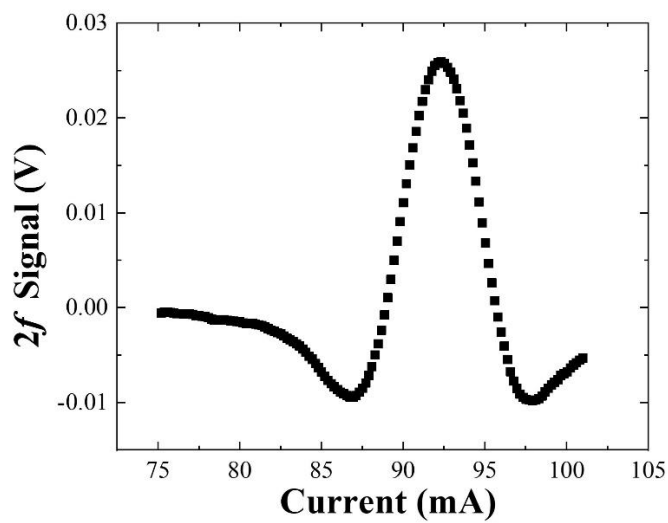

Figure 4 displays the second-harmonic photoacoustic signal of acetylene.

In summary, this manuscript briefly introduces the lithium niobate enhanced laser photoacoustic spectroscopy technique for gas detection, a significant shift from the quartz crystal tuning fork-based photoacoustic spectroscopy systems prevalent over the past 20 years. This new system offers three advantages: 1) Lithium niobate's piezoelectric conversion ability is substantially higher than that of quartz crystals. 2) Lithium niobate crystals have a higher Curie temperature and better temperature stability, enabling a broader application range compared to quartz crystals. 3) Lithium niobate crystals are known as one of the most versatile materials in optics, and based on these, integrated on-chip miniature optical sensors can be fabricated.

Conflict of interest

The authors declare that there are no conflicts of interest.

Acknowledgment

This work is supported by the National Natural Science Foundation of China (12174156, 12174155, 62105125, 62005105, 62075088, 62175137), Natural Science Foundation of Guangdong Province (2020B1515020024, 2019A1515011380).